\documentclass[conference, 10pt]{IEEEtran}
\IEEEoverridecommandlockouts 
\usepackage{cite}
\usepackage{amsmath,amssymb,amsfonts}
\usepackage{algorithmic}
\usepackage{graphicx}
\usepackage{subcaption}
\usepackage{xcolor,multicol,multirow}
\usepackage{enumerate,textcomp,array,enumitem}
\usepackage{threeparttable,url}
\usepackage{dblfloatfix}
\usepackage{flushend}
\usepackage{setspace}
\usepackage{soul}
\usepackage[running,displaymath]{lineno}
\usepackage{etoolbox}
\usepackage{subcaption} 

\def\footnoterule{\relax\if@IEEEenableoneshotfootnoterule
\kern-5pt
\hbox to \columnwidth{\hfill\vrule width 0.5\columnwidth height 0.4pt\hfill}
\kern4.6pt
\global\@IEEEenableoneshotfootnoterulefalse
\else
\relax
\fi}

\usepackage{placeins}
\usepackage{comment}

\def\BibTeX{{\rm B\kern-.05em{\sc i\kern-.025em b}\kern-.08em
    T\kern-.1667em\lower.7ex\hbox{E}\kern-.125emX}}  

\begin{document}
\title{AGI: Agility Index for Power Systems Dynamic Performance, Inspired by Reynolds Number}
\author{
\IEEEauthorblockN{Maria Vabson$^*$, Amir Sajadi$^\ddagger$, Bri-Mathias Hodge$^\dagger$$^\ddagger$}\
\IEEEauthorblockA{$^*$Deptartment of Mechanical Engineering, $^\dagger$Department of Electrical, Computer, and Energy Engineering,}
\IEEEauthorblockA{$^\ddagger$Renewable and Sustainable Energy Institute}
\IEEEauthorblockA{ University of Colorado Boulder, Boulder, CO, USA}

\{maria.vabson, amir.sajadi, brimathias.hodge\}@colorado.edu}

\maketitle
\begin{abstract}
Multiple established metrics and indices exist to quantify the dynamic performance of power systems. With the growth of inverter based resources, these mostly conventional metrics and indices may not encompass all important elements and variables that hold information about the grid performance, especially the damping properties. This paper proposes the Agility Index (AGI), which is a Reynolds number-inspired metric for power system dynamics relating the ratio of inertia to damping. With the current trends towards high-power inverter assets and reduced-inertia grids, explicit attention is necessary in monitoring inertia and damping together. This paper shares insights the AGI presents about the dynamic performance of power systems.
\end{abstract}


\section{Introduction}

Maintaining grid reliability and stability is a pressing concern for power system planners and operators as the number of inverter-based resources (IBRs) continues to grow in the grid. 
This transition often supplants synchronous generators (SGs) and, subsequently, changes the grid's state of inertia and damping. 
The IBR dynamics are constituted by their control software where their response pattern can be modified to produce variable inertia and damping values \cite{intertianrel,syncAmir}. Contrary to IBRs, SGs have electromechanical coupling with the grid and their dynamic performance is restricted to their physical construct, namely their inertia and damping, which are inherent to their design \cite{newcontrol}. This sharp contrast in the design of IBRs vs. SGs and, therefore, their operational principles has fundamental ramifications on power grid dynamics and how healthy performance should be evaluated, especially during operations with a preponderance of IBRs.

From the physics perspective, inertia is a quality of an object to resist changes in its motion, and its momentum quantifies this ability. In power systems, the moment of inertia, most colloquially and henceforth in this paper referred to as "inertia", quantifies a generator's ability to resist changes in frequency \cite{intertianrel}. In SGs, when there is a disturbance, stored kinetic energy (KE) in the spinning rotor can briefly sustain power output until the control processes are activated or the disturbance is removed. This gives the response systems in the grid a moment to detect the change in frequency and properly respond. Now, there is growing interest in damping because of its role in influencing stability and synchronization in the system \cite{syncAmir}. The damping affects the system's ability to dissipate oscillations and return to an equilibrium, analogous to a mechanical system where higher damping capacity allows for more effective KE dissipation through the network \cite{syncAmir}. The interplay between inertia and damping is critical as simply increasing inertia to compensate for increased IBRs without considering damping may be ineffective \cite{del2021effects}. 

With more IBRs coming online, the minimum amount of inertia necessary to maintain frequency stability is under question. There are two main control technologies for IBRs; grid-following (GFL) and grid-forming (GFM) control technology. GFL can be viewed as a controllable negative load, because they only control active and reactive power injection as a current source while simply tracking the voltage signal at its point of interconnection (POI) \cite{syncAmir}. On the other hand, GFMs directly have inertial and damping impacts on grid frequency behavior as they form and regulate their voltage waveform at the POI without reliance on the grid's condition as a voltage source. However, the inertia and damping coefficients of GFMs are much smaller than those of an SG. Nevertheless, from the classical dynamics perspective, an object's smoothness of motion is a function of its relative values of inertia and damping, since their individual values could be insufficient to adequately describe its agility. This concept forms the basis for the motivation behind this paper. 

This paper proposes an Agility Index (AGI) for modern power systems, inspired by the Reynolds number (Re) in fluid dynamics, which relates inertial and opposing forces. This allows for the characterization of a power grid's dynamics in a similar manner. The AGI measures the behavior a power system and can help estimate the adjustments needed to improve transients without having to conduct computationally-intensive time-domain dynamic simulations, and using only the inertia and damping characteristics of generators. For validation, the AGI was compared against the standard metrics and indices that are common in evaluating the dynamic performance of power systems. The results suggest that the AGI value seems most closely related to the Hertz-sec metric (HS) and settling time (ST). 

\section{Theory of the Proposed Idea}

\subsection{Reynolds Number}
Re is a dimensionless number that describes fluid flow. It established a relationship to understand and predict the behavior of fluids in various flow situations \cite{saldana2024reynolds}. Re is calculated by taking the ratio of the inertial force that sustains motion to the viscous force, which is the internal resistance that opposes flow, as shown in Equation \eqref{eqn1}. It is described by the characteristics of the system, like a channel or pipe, and the fluid that is flowing. For simple pipe flow, the inertial forces are defined by the density ($\rho$), velocity ($v$), and the characteristic length, such as the pipe diameter ($L$), which is analogous to momentum. The viscous term is defined by the dynamic viscosity of the fluid ($\mu$) which represents the fluid's resistance to flow \cite{saldana2024reynolds}. 

\begin{equation} \label{eqn1}
 \text{Re} = \frac{\text{Inertial  Force}}{\text{Viscous  Force}}  = \frac{\rho v L}{\mu}
\end{equation}

Equation \eqref{eqn1} highlights the practicality of being able to relate the two main forces describing the flow, and simplifying them to measurable parameters. As inertial forces increase, there is more energy in the fluid, making it more chaotic and turbulent. With less velocity or smaller pipe dimensions, there is less energy at a given cross-section. Generally, simple pipe flow with a Re below $2,000$ is laminar, meaning damping forces dominate. Conversely, a Re above $4,000$ describes a turbulent flow area where inertial forces dominate. In between, the transition region can be described by a combination of both flows \cite{txt1}. While these ranges are not true for all flow regimes, it is a beneficial metric to characterize flow without solving the more difficult associated partial differential equations, i.e. the Navier-Stokes equations \cite{saldana2024reynolds}. The Re also helps reduce the Navier-Stokes equation into a solvable format, like the Euler equations, when the Re is high and the flow is assumed to be inviscid \cite{NASA}.

There have been numerous developments in the field expanding the applications of Re to fluids with complex viscosities and behaviors like polymers and gels by creating more detailed expressions of Re \cite{saldana2024reynolds}. We posit that there are possible applications of Re to a different class of dynamical systems such as electric power grids, due to the analogous concepts observed in their behavior and complexities. Inertia can be explained similarly in both fields as the force describing continuing motion and acceleration. The damping in power systems is similar to viscosity in fluids in that it contributes to dissipating and smoothing out transient motions. 

\subsection{Reynolds Number for Power Systems: AGI}
The classical model of power-frequency dynamics in power systems is described by the swing equation, as in Equation \eqref{swing1} \cite{sauer2006power}: 

\begin{equation} \label{swing1}
M \frac{d\omega}{dt} =  P_m -  P_e - D (\omega-\omega_s) 
\end{equation} 
where $\omega$ is the angular frequency and $\omega_s$ is the synchronizing angular frequency, e.g., $377$ in North America. In this model, $P_m$ and $P_e$ are the input mechanical power that drives the generator shaft and electrical power injected into the grid, respectively. $M$ and $D$ are the angular momentum and damping coefficient respectively, $\{M,D\}>0$. In the literature, power system dynamics are widely interpreted through the lens of inertia, represented by $M$. Often, focusing on the reduced inertia aspect of modern power grids overlooks the significance of damping, represented by $D$, as an equally contributing element in helping the frequency reach a steady state, as demonstrated in \cite{walth2025lunar}.

To define AGI, the Re analog for power systems, Equation \eqref{swing1} can be linearized and rearranged as \cite{machowski2020power}: 
\begin{equation} \label{swing2}
 M \frac{{d\Delta \omega}}{dt} + D \Delta \omega = P_a
\end{equation} 
where $P_a=P_m-P_e$ and $\Delta$ is linear operator. This is a first-order differential equation for $\Delta \omega$. Solving this equation gives the angular frequency deviations as
\begin{equation} \label{swing3}
\Delta \omega(t) =  \epsilon \Big(1-  e ^{-\frac{t}{\tau}} \Big)
\end{equation} 
where $\epsilon=\frac{P_a}{D}$ is the steady state value of $\Delta \omega$ and $t$ is time. In equation \eqref{swing3}, $\tau=\frac{M}{D}$ is the time decay constant which
is dependent on the generator's angular momentum ($M$) and its damping coefficient ($D$). 
This naturally coincides with the structure of the Re, defining AGI as shown in Equation \eqref{ReP}: 
\begin{equation} \label{ReP}
 AGI =\frac{\text{Angular Momentum}}{\text{Damping Coefficient}} =  \frac{M}{D}  
\end{equation} 
The value of $M$ is computed using the inertia time constant, $H$, which is defined as the relationship between the stored KE and machine rating, $S_n$, as $M = \frac{2H S_n}{\omega_s}$. The units of $M$ are kg.m$^2$.s$^{-1}$ \cite{machowski2020power}, and replacing 1N=1kg.m.s$^{-2}$ conveniently gives its alternative units of N.m.s. The units of $D$ are N.m as it represents all that contributes to effective damping in the form of  counteracting torque \cite{machowski2020power}. 
Based on the units of $M$ and $D$, the AGI has units of time in second, $s$, which is consistent with its physical nature as a time constant, as shown in Equation \eqref{swing3}.

\begin{figure}[!h]
    \centering
    \includegraphics[width=\linewidth]{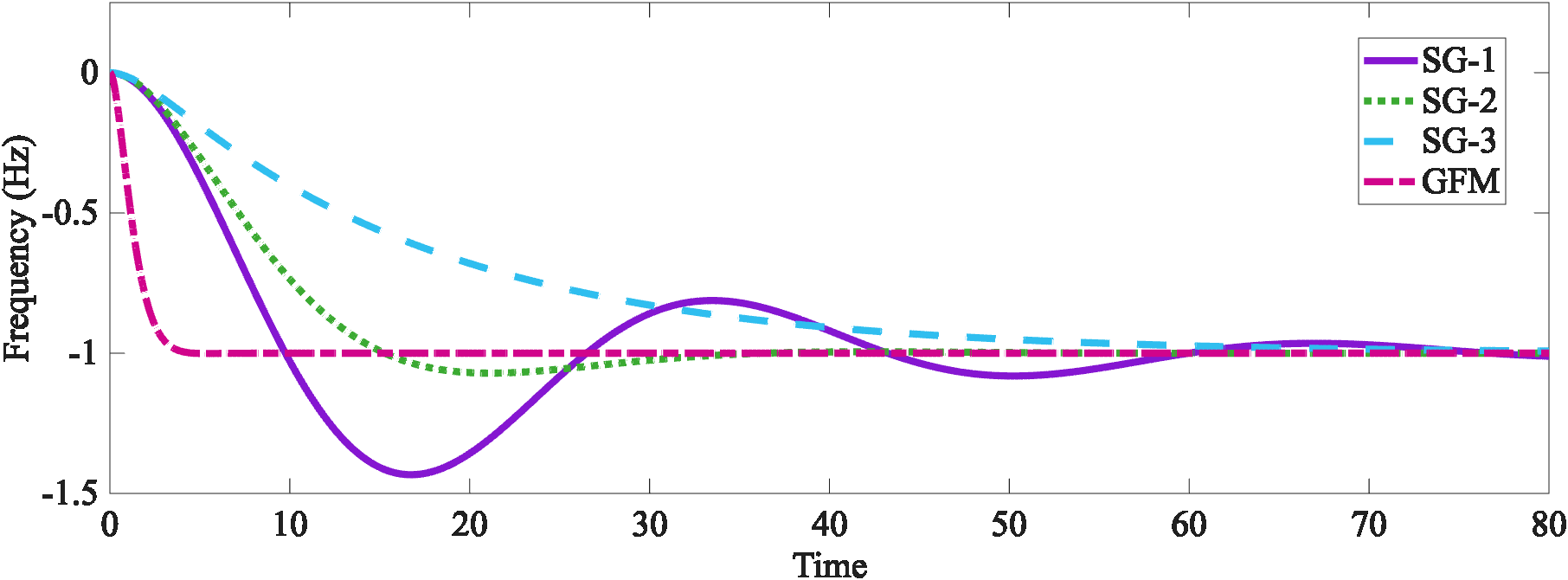}
    \caption{Comparison of step responses from Table \ref{tab:agicalc}.}
    \label{fig:stepresponses}
\end{figure}

Examples of AGI for a set of SG and GFM are calculated in Table \ref{tab:agicalc} and their corresponding time-domain frequency responses are displayed in Figure \ref{fig:stepresponses}. A clear distinction can be noted between the first-order-like response of the GFM with the AGI below $1$ vs. the second-order response of SGs with AGI values above $2$. Among the SGs, SG-1 which has the largest AGI value exhibits the most oscillations and the longest decay, while SG-2 with a smaller AGI returns to an equilibrium faster with a smoother trajectory. SG-3 decays slowly like the other SGs but the most smoothly of the SGs. The GFM case which has the smallest AGI of all the cases considered settles the quickest without oscillations. Based on this observation and the physical construct of SGs (where most commonly $\frac{D}{M}<0.5$ \cite{sauer2006power,machowski2020power}) and GFM (where most commonly $\frac{D}{M}>1$ \cite{trujillo11328929,syncAmir,Wallace_interactive}), we suggest that the values of AGI under $1$ indicate a more laminar flow, consistent with the characteristics of GFM-dominant systems. Whereas, the values of AGI over $2$ represents the more turbulent flow like the Re represents for its use, consistent with the characteristics of SG-dominant systems. The AGI range between $1$ and $2$ may suggest behavior of transitional flow, which would be a mixture of GFM and SG characteristics as exhibited by SG-3 which has an AGI value of $1.5$. 
\begin{table}[!h] 
    \centering
        \caption{AGI values  for S=1 MVA and $\omega=377$, from \cite{trujillo11328929}}
    \renewcommand{\arraystretch}{1.1}
    \begin{tabular}{l|c|c|c|c|c}
    \hline
       Type & H (s)& M (Nms) & D (pu) & D (Nm) & AGI (s)  \\
    \hline
     SG-1   &  5 &  26,525   &  0.10 &  2,652.5 & 10 \\ 
    \hline   
    SG-2   &  5 & 26,525   &  0.25 &  6,631   & 4.0 \\ 
    \hline 
    SG-3   &  5 & 26,525   &  0.67 &  17,683   & 1.5 \\ 
    \hline  
       GFM  & 0.1 & 530.5  &  10 &  5,305 &  0.1  \\ 
    \hline
    \end{tabular}
    \label{tab:agicalc}
\end{table}
\section{Numerical Simulation}

\subsection{Simulation Description}
\begin{figure}[h]%
    \centering
    {{\includegraphics[width=0.49\textwidth]{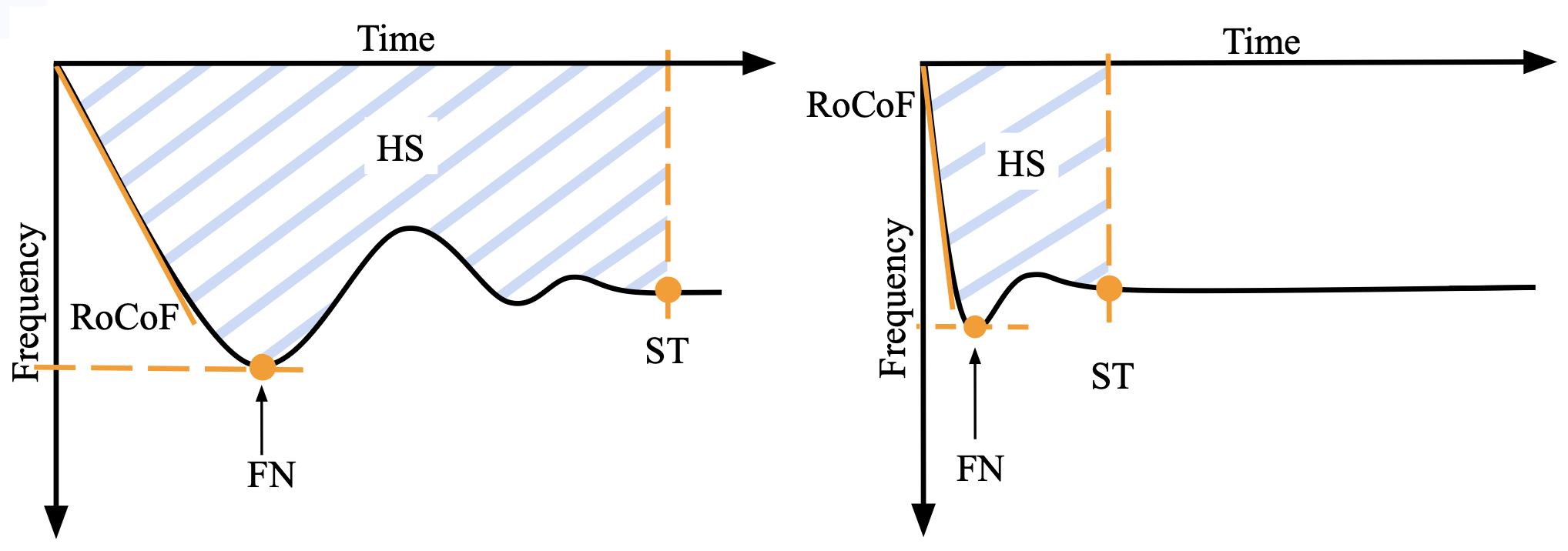} }}%
    \caption{The transient responses of a SG (left),  GFM (right)}
    \label{fig:metricvis}%
\end{figure}

We used the 9-bus, 3-generator IEEE test system \cite{powerdata} for numerical simulations. A perturbation of fixed magnitude was introduced on a generator in the simulation of 1000 cases with random allocations of inertia and damping coefficients for the generators in the system in each case. Despite the randomness, the $\frac{D}{M}$ ratio values were restricted to 0.025 (on the lower bound) and 100 (on the upper bound) to emulate realistic set of inertia and damping values of SG and GFM \cite{sauer2006power,trujillo11328929}. For each iteration, the dynamic performance of the perturbed generator was quantified using the following metrics to compare with the AGI: Rate of Change of Frequency (RoCoF), Frequency Nadir (FN), Settling Time (ST), the Center of Inertia (COI) and the Hertz-sec metric (HS). These metrics are commonly used in academic and industrial studies and describe certain characteristic of frequency response, as visualized in Fig. \ref{fig:metricvis}. Additionally, the ST/RoCoF ratio is also considered for a more comprehensive comparison. The definitions of these metrics are discussed in what follows. 

\subsection{Dynamic Performance Metrics}
Frequency regulation is typically assessed by RoCoF and FN. The RoCoF is the derivative of the frequency with respect to time and has traditionally quantified the robustness of the grid's ability to withstand disturbances \cite{rocof}. The FN is the minimum frequency reached after a disturbance \cite{newcontrol}. In full SG systems, high RoCoF meant low FN, but this no longer holds true for heterogeneous systems, demanding better metrics and analysis of frequency response \cite{Wallace_interactive}. The ST is the time it takes for a signal to return to its normal range after a disturbance \cite{eto2011use}. In addition, the ST/RoCoF is calculated as a proxy of the frequency settling behavior relative to the rate of initial decline as it encompasses how the system responds initially with RoCoF and finally settles in one metric. The COI is another relevant metric and is defined as the inertial center of all the generators of interest \cite{rocof}. COI is calculated as shown in Equation \eqref{coieqn} with essentially the weighted average of the generator's inertia constant ($H$) relative to the power generated ($P_g$) \cite{kundur1994power}. This metric helps represent the behavior of the grid, but does not represent actual physical locations. 
\begin{equation} \label{coieqn}
COI = \frac{\sum (H * P_g)}{\sum P_g}
\end{equation}

\noindent Finally, the frequency response in power grids with heterogeneous generation characteristics can also be quantified using HS \cite{syncAmir}. It represents frequency transients by taking the integral of the change in frequency from pre-perturbation frequency $f_0$ to post-perturbation frequency $f(t)$ from the time of the perturbation $t_0$ until the frequency settles, as described in Equation \eqref{HSequation} and shown in Fig. \ref{fig:metricvis} as the striped area until the ST is reached. 
\begin{equation}\label{HSequation}
    HS= \int^{ST}_{t_0} |f_0 - f(t)|dt
\end{equation}
The lower the HS, the more GFM-like the response.

\begin{figure*}[!]%
    \centering
    {\includegraphics[width=\textwidth]{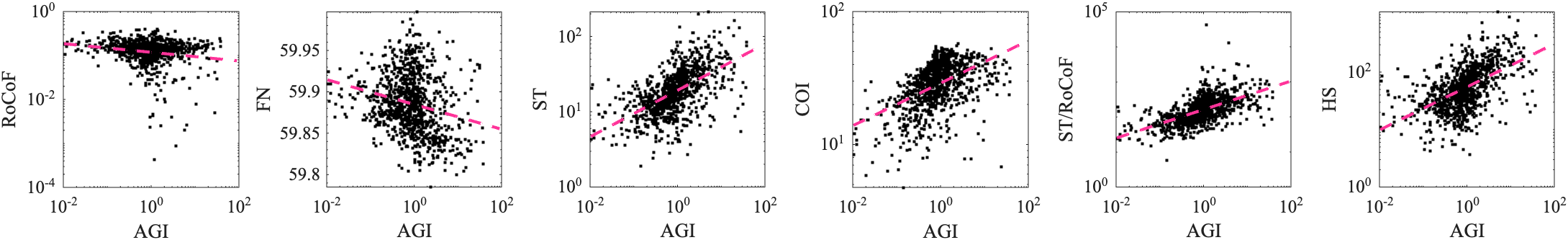} }
    \caption{The log-log relationships of the metrics of interest compared to AGI}
    \label{fig:withcutoffnot normalized}
\end{figure*}

\section{Results and Discussion}

\subsection{Comparative Performance}
The AGI was compared to the dynamic performance metrics, listed in the previous section, to analyze its utility and efficacy. The log-log relationships were used to normalize the metrics for a more interpretable visualization in Fig. \ref{fig:withcutoffnot normalized}. The linear log-log fit is plotted on top of the data as the dotted pink line, so data that most closely follows this is better correlated. From the visual inspection, while no perfectly linear relationship emerged, the strongest relationship was seen between AGI and ST, ST/RoCoF, and HS, as well as COI. We suggest that ST and HS correlation to AGI is justified because the lower ST is typically seen in systems with lower inertia and higher damping (which naturally forms the conditions for lower AGI), and HS is typically smaller in those more GFM dominant systems with the same attributes too. Similarly, as AGI increases those metrics would as well. This is most interesting as ST and HS implicitly considers damping but AGI does it more explicitly and can be calculated without the needs to evaluate the time-domain frequency response. Further, the ST/RoCoF correlation with AGI makes sense building on the correlation of ST and AGI, along with RoCoF being singularly related to inertia, so the positive correlation tracks. Finally, the positive correlation of COI and AGI makes sense since the numerator of AGI is inertia. 

\subsection{Statistical Correlation}
To further understand the correlations, the root mean square error (RMSE), coefficient of determination (R$^2$), and Pearson's correlation coefficient ($\rho$) values for the relationships between the logarithmic values of both the AGI and the other metrics considered were calculated and are shown in Table \ref{tab:error}. 
The RMSE measures the average distance between the predicted and actual values, estimating how well the model can predict the true values \cite{chicco2021coefficient}. The closer the RMSE to zero, the predicted values are exactly linear on the log scale. 
Another metric used is the $R^2$ value which measures the amount of variation in the response variable that is explained by the independent variable; the values closest to $1$ match the linear fit \cite{chicco2021coefficient}. 
Lastly, $\rho$ is a global error measure metric, with larger values indicating higher linear correlation on the log scale \cite{zhang2013metrics}.

\begin{table}[h!]
\caption{The RMSE, $R^2$, and r between logarithmic values of AGI and the metrics of interest \label{tab:error}}
\centering
\renewcommand{\arraystretch}{1.35}
\begin{tabular}{c|c|c|c|c|c|c}
\hline
Value & RoCoF & FN & ST & COI & $\frac{ST}{RoCoF}$ &  HS\\
\hline
RMSE  & 0.27  & 0.0  & 0.24 & 0.15 & 0.30 & 0.32 \\  \hline
$R^2$ & 0.04  & 0.06 & 0.33 & 0.25 & 0.35 & 0.29 \\ \hline
$\rho$ & -0.19 & -0.24 & 0.57 & 0.50 & 0.59 & 0.54 \\ \hline
\end{tabular}
\end{table}

In testing various relationships to correlate the AGI value to the performance of the model in a meaningful way, the $R^2$ and $\rho$ support that the strongest relationships to be with the ST, ST/RoCoF, the HS metric, and also the COI. The RMSE is skewed higher for the metrics that span more orders of magnitude, not fully representing the relative amount of error between metrics. The correlation with ST/RoCoF and HS is similar to ST, and even slightly higher in terms of $R^2$ and $\rho$ for ST/RoCoF. However, the statistical similarity between AGI and metrics like RoCoF and FN were minimal as these may not have as unique behavior in SGs vs. GFMs, yet the slight negative correlation does support how more SG dominant systems may see lower RoCoFs and FNs. The positive correlation of COI and AGI follows with their relationship with inertia and all of these results corroborate with the previous results. 

\subsection{Probability Density Estimates}

\begin{figure}[h!]%
    \centering
    {\includegraphics[width=\linewidth]{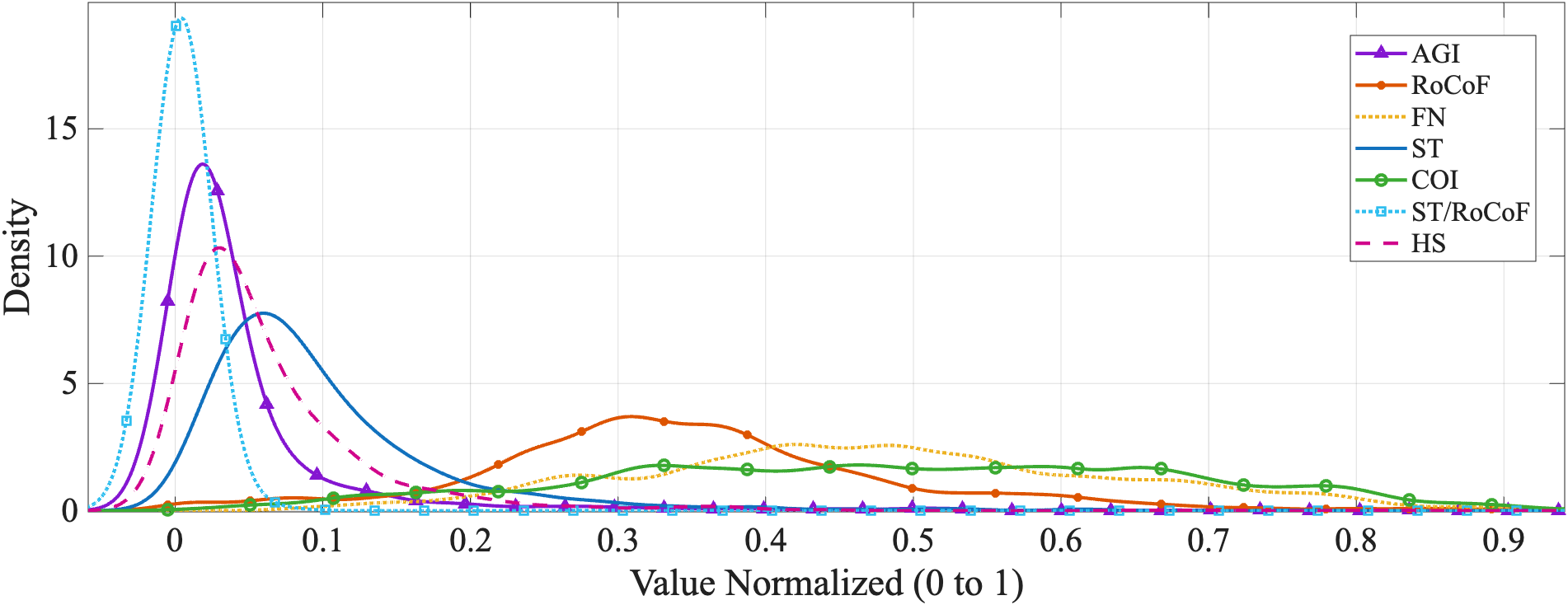}}%
    \caption{The PDEs of the normalized metrics.}%
    \label{fig:normpdes}%
\end{figure}

\begin{figure}[h!]%
    \centering
    {\includegraphics[width=\linewidth]{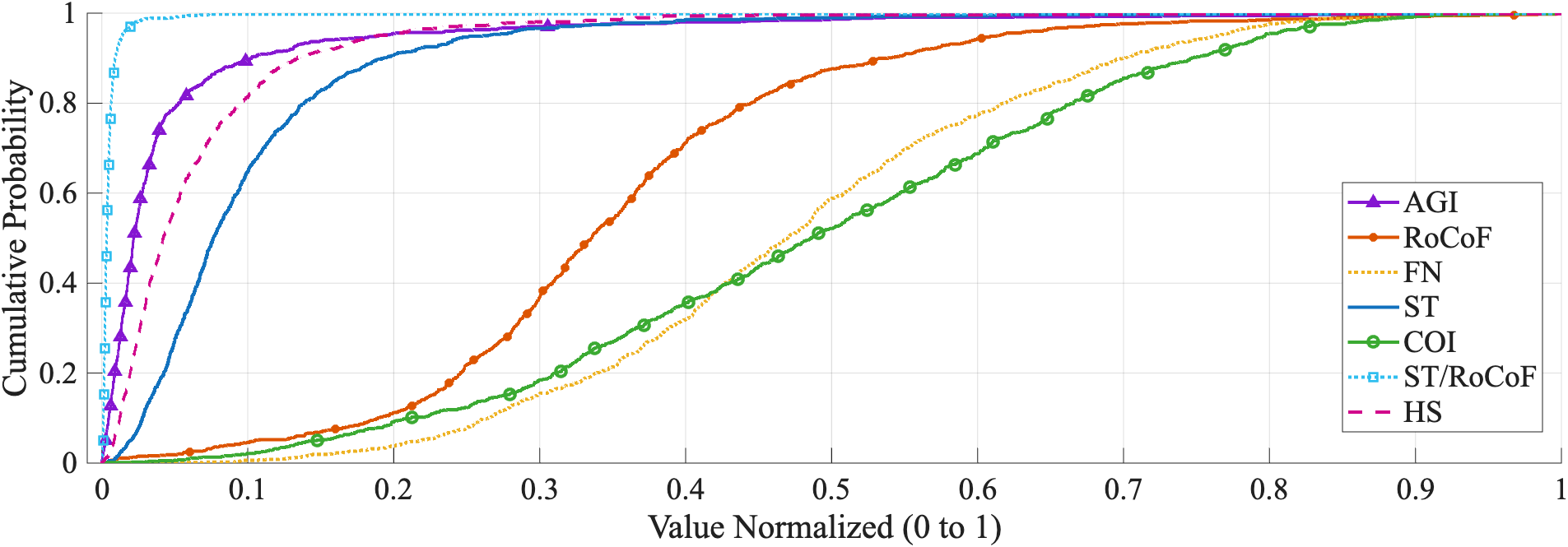}}%
    
    \caption{The CDFs of the normalized metrics.}
    \label{fig:normcdf}%
\end{figure}

To further compare the performance of AGI vs. all other metrics, probability density estimates (PDEs) and cumulative density functions (CDFs) of the data are plotted in Fig. \ref{fig:normpdes} and Fig. \ref{fig:normcdf} respectively. 
To do so, each respective metric was first normalized preserving the distance between the data points and scaled between $0$ and $1$ as shown by Equation \eqref{norm2} to enable fair comparison of the metrics' PDEs and CDFs. 
\begin{equation} \label{norm2}
x_{\{0\ to \ 1\}}= \frac{x_i - min(x)}{max(x) - min(x)} 
\end{equation}
The kernel smoothing Matlab function is used to create the PDEs which approximates the true probability density functions. 

The results suggest that AGI is most similar in distribution to the HS, followed closely by the ST, and then ST/RoCoF in terms of the skewed nature in of PDEs in Fig. \ref{fig:normpdes}. Similarly, the CDF of AGI most closely resembles HS and then ST.These correspond with previous results seen in Fig. \ref{fig:withcutoffnot normalized} and Table \ref{tab:error}. The ST/RoCoF exhibits a similar overall distribution shape to AGI, except with a much longer tail. While ST/RoCoF does not have a standard use or definition, its utility could be of interest for further exploration. The distributidon of COI differs significantly from the AGI compared to their seemingly closer correlation in the previous analysis and statistics, but holds as COI is only a measure of inertia. These results further reveal the applicability of AGI in containing similar information to the post disturbance calculated response metrics, primarily the HS and ST with information known prior to having to conduct any simulation or analyze the frequency response. 

\subsection{Discussion}

We suggest that the results here indicate that the AGI can provide virtually the same output of the state of dynamics of the system without having to run simulations or evaluate the time-domain frequency response. The AGI may be utilized as an estimate of a system's stability to inform adjustments and optimization of inertia and damping, and may be expanded to a notion of a reasonable range to maintain a system. This is a major benefit, as inertia and damping can be adjusted in near real-time for GFMs based on the AGI value. It helps characterize systems with low AGI values and adequate damping or more turbulent systems that take longer to settle and oscillate more that have higher AGI. 

With the simplicity of the initial network, this is only a starting point for further analysis into the reliability of the AGI metric for the dynamic characteristics of large-scale and complex power grids. Deeper analysis needs to be done to create a more comprehensive reference guideline based on calculated values for different systems similar to the Moody diagram used in fluid dynamics which relates Re to varying levels of friction and roughness to determine the regions of laminar or turbulent flow in different environments \cite{fluidmechanicsfrank}. 

\section{Conclusion}
With more IBRs and the grid trending to lower inertia levels, it is critical to understand how stability may be affected, possibly by keeping track of both the inertia and the damping. This paper proposed a new agility index (AGI) for assessing power grid dynamics based on inertia and damping coefficients, similar to the fluid dynamics concept of Reynolds number (Re). This index enables an assessment of the system's behavior without necessarily needing to conduct simulations or having to analyze the frequency response. The initial results show a positive correlation between the proposed AGI with Hertz-second (HS) and settling time (ST), suggesting a plausible path to capture insight into the the system's expected frequency response.
\bibliographystyle{IEEEtran}
\bibliography{sources}

\end{document}